\documentclass[preprint2]{aastex}
\usepackage{lineno}

\newcommand{\mathsym}[1]{{}}
\newcommand{\unicode}[1]{{}}
 \usepackage{graphics}
 \usepackage{graphicx}
 \usepackage{epsfig}

\usepackage{amssymb}
\usepackage{amsmath}
\usepackage{aas_macros}

\shorttitle{Analysis of Equilibrium Points\dots Four Body Problem}
\shortauthors{Reena Kumari \and Badam Singh Kushvah}

\begin{document}


\title{Stability Regions of Equilibrium Points in Restricted Four-Body Problem with Oblateness Effects}


\author{Reena Kumari \and Badam Singh Kushvah}
\affil{Department of   Applied  Mathematics,\\
Indian School of Mines, Dhanbad - 826004, Jharkhand,India}
%
\email{reena.ism@gmail.com; bskush@gmail.com} 

\begin{abstract}
In this paper, we extend the basic model of the restricted four-body problem introducing two bigger dominant primaries $m_1$ and $m_2$ as oblate spheroids when masses of the two primary bodies ($m_2$ and $m_3$) are equal. The aim of this study is to investigate the use of zero velocity surfaces and the Poincar\'{e} surfaces of section to determine the possible allowed boundary regions and the stability orbit of the equilibrium points. According to different values of Jacobi constant $C$, we can determine boundary region where the particle can move in possible permitted zones. The stability regions of the equilibrium points expanded due to presence of oblateness coefficient and various values of $C$, whereas for certain range of $t$ ($100 \le t \le 200$), orbits form a shape of cote's spiral. For different values of oblateness parameters $A_1~ (0<A_1<1)$ and $A_2~ (0<A_2<1)$, we obtain two collinear and six non-collinear equilibrium points. The non-collinear equilibrium points are stable when the mass parameter $\mu$ lies in the interval ($0.0190637,~0.647603$). However, basins of attraction are constructed with the help of Newton Raphson method to demonstrate the convergence as well as divergence region of the equilibrium points. The nature of basins of attraction of the equilibrium points are less effected in presence and absence of oblateness coefficients $A_1$ and $A_2$ respectively in the proposed model.
\end{abstract}


\section{Introduction}
\label{sec:Int}
To study the motion of celestial bodies, restricted four-body problem is one of the  important problem in the dynamical system. An application of the restricted four-body problem is illustrated in the general behavior of the synchronous orbit in presence of the Moon as well as the Sun whereas coupled restricted three-body problem is one of the example of restricted four-body problem. The problem is restricted in the sense that one of the masses is taken to be small, that the gravitational effect on the other masses by the fourth mass is negligible. The smaller body is known as infinitesimal mass (body) and remaining three finite massive bodies called primaries.

The classical restricted four-body problem may be generalized to include different types of effect such as oblateness coefficient, radiation pressure force, Pyonting-Robertson drag etc. Various authors have studied the restricted four-body problem and examined the existence of equilibrium points such as  \cite{Hadjidemetriou1980CeMec..21...63H}, \cite{Michalodimitrakis1981Ap&SS..75..289M},  \cite{Kalvouridis2007P&SS...55..475K}  and \cite{Papadakis2007P&SS...55.1368P}. Further,  \cite{Baltagiannis2011Ap&SS.336..357B} discussed the equilibrium points and their stability in the restricted four-body problem.

On the other hand, in recent years many perturbing forces, such as oblateness, radiation forces of the primaries, Coriolis and centrifugal force, variation of the masses of the primaries etc. have been included in the study of restricted three-body problem (RTBP). The RTBP with oblate effect has been studied by many investigators such as  \cite{Sharma1975CeMec..12..189S}, \cite{Abouelmagd2012Ap&SS.341..331A}, \cite{Khanna1999IJPAM..30..721K},  \cite{Douskos2011Ap&SS.333...79D} etc.

Determination of the stability regions of the infinitesimal body was introduced by \cite{Poincare1892QB351.P75......} during the study of periodic orbit of the system. This is very good technique to study the nature of trajectory of an infinitesimal body and also known as surface of section method. Apart from that this method was used by \cite{Winter2000P&SS...48...23W} and \cite{Kumari2013Ap&SS.344..347K} to describe the location and stability of the equilibrium points in the restricted three and four-body problem respectively.

Here, we extend the basic model of restricted four-body problem by considering the dominant primary $m_1$ and $m_2$ as oblateness body respectively. Our goal in this paper is to study the effect of oblateness coefficient on the motion of an infinitesimal body in the force field of massive bodies. We also determine and present basins of attraction
for the equilibrium points (attractors) of the problem created by Newton Raphson method for their numerical computation at sample values of the oblateness coefficient parameter. The set of initial approximation $(x, y)$ which leads to a particular equilibrium
point, constitutes a convergence (or attracting) or divergence region. \cite{Douskos2010Ap&SS.326..263D}  and \cite{Croustalloudi2007P&SS...55...53C} presented a similar study of the basins of attraction in the $xy$-plane for the equilibrium points  of Hill's problem with radiation and oblateness in restricted three body problem and of a ring problem of $n+1$ bodies.

The Poincar\'{e} surface of section of the proposed model is obtained with the help of the Event Locator Method. We have used  Mathematica$^{\textregistered}$ \cite{wolfram2003mathematica} software package for numerical and algebraic computation of non-linear ordinary differential equations.

This paper is organized as: we write the equations of motion and find the Jacobi integral of the system in section (\ref{sec:veqn}). In section (\ref{sec:zvc}), we describe the zero velocity surfaces whereas in section (\ref{sec:equn}) we determine equilibrium points. The stability of the equilibrium points is examined in section (\ref{sec:lsncp}) and (\ref{sec:pss}) whereas in section (\ref{sec:basin}) we present interesting
basins of attraction created by Newton Raphson method applied for the solution of
the equations whose roots provide the locations of the equilibrium points. Finally, section (\ref{sec:con}) includes the discussion and conclusion of the paper.

\section{Equations of motion}
\label{sec:veqn}
\begin{figure}[t]
 \begin{center}
  \plotone{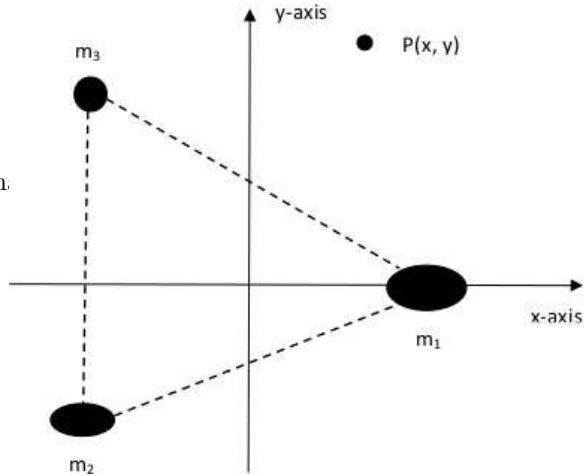}
\caption{Geometry of the problem.\label{fig:fig1}}
 \end{center}
\end{figure}
In this problem, we suppose that the motion of an infinitesimal mass $(m)$ is governed by the gravitational force of the oblate spheroid $ m_1$, $ m_2$  and third body $m_3$ with $m_1 > m_2\ge m_3$ (\ref{fig:fig1}). The oblateness factor of the primaries ($m_1, m_2$) are also taking into account. It is assumed that the influence of infinitesimal mass on the motion of primaries moving under their mutual gravitational attraction is negligible. We normalize the units with the supposition such that the sum of the masses and separation between the primaries both be unity and unit of time is taken as the time period of rotating frame moving with the mean motion $(n)$. Hence, we have $G(m_1+m_2+m_3)=1$. Let the co-ordinates of infinitesimal mass be $(x, y)$ and masses $m_1$, $m_2$  and $m_3$ are $(\sqrt{3}\mu, 0)$, $(-\frac{\sqrt{3}}{2}(1-2\mu), -\frac{1}{2})$ and $(-\frac{\sqrt{3}}{2}(1-2\mu),\frac{1}{2})$ respectively, relative to rotating frame $Oxyz$, where $\mu=\frac{m_2}{m_1+m_2+m_3}=\frac{m_3}{m_1+m_2+m_3}$ is the mass parameter and we assume that $\mu=0.2$. The perturbed mean motion $n=\sqrt{1+\frac{3}{2}(A_1+A_2)}$, where $A_i=\frac{R^{2}_{e_i}-R^{2}_{p_i}}{5R^{2}}, i=1,2 $ is oblateness coefficient of oblate bodies $m_1$ and $m_2$ respectively with $R_{e_i}$ and $R_{p_i}$ as equatorial and polar radii and $R$ is separation between the primaries.

The equations of motion of the infinitesimal mass in the rotating co-ordinate system is given as 
\begin{eqnarray}
\ddot{x}-2n\dot{y}&=&\Omega_{x}, \label{eq:ux}\\
\ddot{y}+2n\dot{x}&=&\Omega_{y}, \label{eq:vx} 
 \end{eqnarray}where
 \begin{eqnarray}
&&\Omega=\frac{n^{2}(x^{2}+y^{2})}{2}+\frac{(1-2\mu)}{r_1}+\frac{\mu}{r_2}+\frac{\mu}{r_3} \nonumber \\&&+\frac{(1-2\mu){A_1}}{{2}r^{3}_1}+\frac{\mu{A_2}}{{2}r^{3}_2},
 \end{eqnarray}
with
\begin{eqnarray}
&& r_1=\sqrt{(x-\sqrt{3} \mu)^{2}+y^{2}},\nonumber\\&& r_2=\sqrt{\left(x+\frac{\sqrt{3}}{2}(1-2 \mu)\right)^{2}+\left(y-\frac{1}{2}\right)^{2}},\nonumber\\&& r_3=\sqrt{\left(x+\frac{\sqrt{3}}{2}(1-2\mu)\right)^{2}+\left(y+\frac{1}{2}\right)^{2}},\nonumber\\&&  r=\sqrt{x^{2}+y^{2}}.\nonumber
\end{eqnarray}
The suffixes $x$ and $y$ indicate the partial derivatives of $\Omega$ with respect to $x$ and $y$ respectively.
The well known energy integral of the problem given as:
\begin{eqnarray}
&&C=-\dot{x}^{2}-\dot{y}^{2}+2\Omega, \label{eq:ji}
 \end{eqnarray}
 where $C$ is known as Jacobi constant. We observe (from \ref{eq:ji}) that  $2\Omega-C\ge0$. The curves of zero velocity are defined through the expression $2\Omega=C$; such a relation defines a boundary, called Hill's surface, which separates regions where motion is allowed or forbidden.

\begin{figure}[t]
 \begin{center}
  \plotone{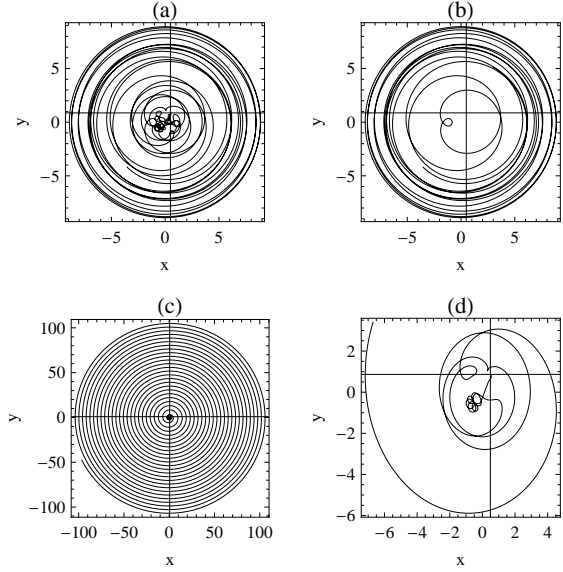}
\caption{Orbits of the restricted four-body problem with and without oblateness effect.\label{fig:fig2}}
 \end{center}
\end{figure}
In Fig.\ref{fig:fig2}, four frames represent the orbit of the infinitesimal body. First two frames show the orbit in absence of oblateness effect whereas last two frames show orbit in presence of oblateness effect. The orbit of the infinitesimal body represents in first frame when $0 \le t \le 200$ whereas second frame when $100 \le t \le 200$. In the second frame, we observed that in absence of oblateness effect, orbit looks like cote's spiral. However, with effect of oblateness, orbit  becomes regular when $0 \le t \le 200$ which is shown in third frame while fourth frame shows the orbit when $0 \le t \le 50$. 
\section{Zero velocity surfaces}
\label{sec:zvc}
Eq.(\ref{eq:ji}) represents a relation between square of velocity and the coordinates of the infinitesimal body in the rotating coordinate system. The Jacobi constant $C$ is determined numerically using initial conditions. Therefore equation (\ref{eq:ji}) determines the boundaries of the regions where the body can move from one allowed region to another one. In particular, if we take velocity of the infinitesimal body equal to zero then surfaces obtained in $xy$-plane known as zero relative velocity surfaces which are given as follows: 
\begin{eqnarray}
 C=2\Omega \label{eq:zvc1}
\end{eqnarray} or
\begin{eqnarray}
&& {n^{2}(x^{2}+y^{2})}+\frac{2(1-2\mu)}{r_1}+\frac{2\mu}{r_2}+\frac{2\mu}{r_3}\nonumber\\&&+\frac{(1-2\mu){A_1}}{r^{3}_1}+\frac{\mu{A_2}}{r^{3}_2}=C. \label{eq:zvc2}
\end{eqnarray}
The above solution gives much information about the possible dynamics at a given Jacobi constant $C$. In particular, if $A_1=A_2=0$ in equation (\ref{eq:zvc2}) we obtain the classical zero velocity surfaces of the system, to study the  behavior of the zero velocity surfaces in the vicinity of the singular point and in the vicinity of the main bodies for increasing and decreasing values of Jacobi constant.
\begin{figure}[h]
 \begin{center}
  \plotone{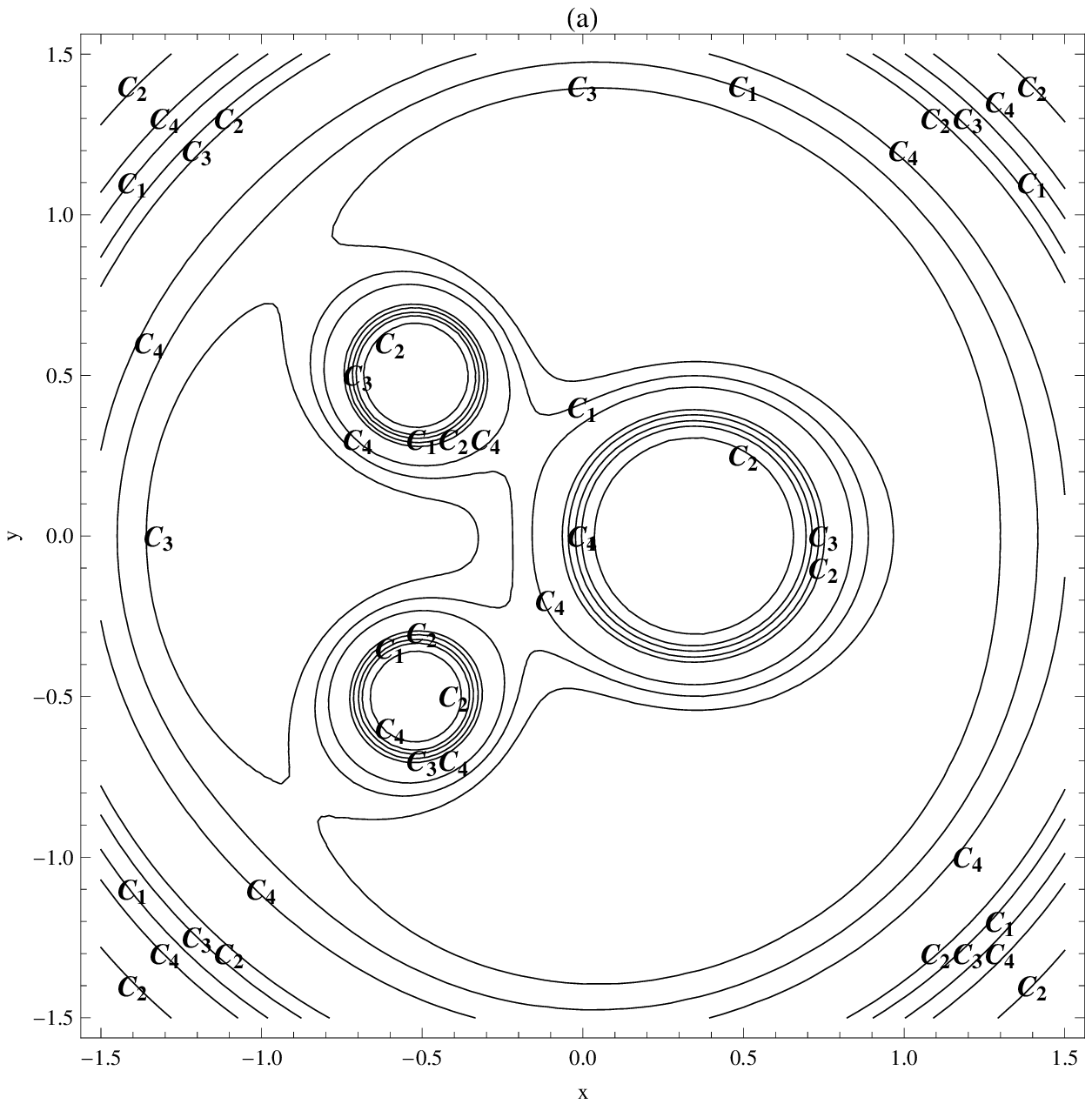}\\ \plotone{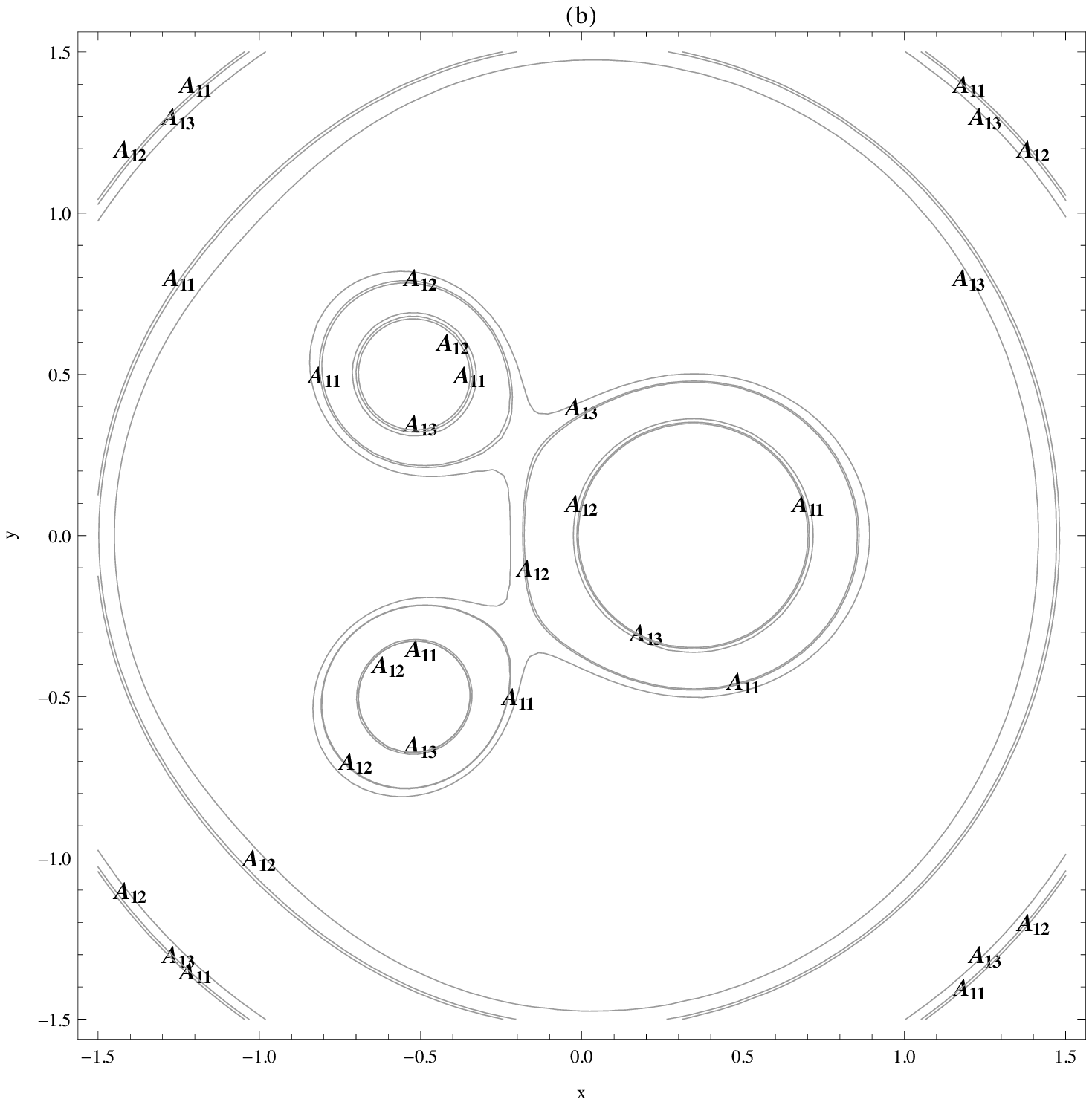}
\caption{Zero velocity curves for (a) $ C_1=3.5, C_2=2.5, C_3=1.7$ and $C_4=1.0$ and (b) $A_{11} (A_1=0.0, A_2=0.0), A_{12} (A_1=0.0025, A_2=0.0025)$, and $A_{13} (A_1=0.005,A_2=0.005)$.\label{fig:fig3}}
 \end{center}
\end{figure}

In Fig. \ref{fig:fig3}, frame (a) shows zero velocity curves (ZVC) for different values of Jacobi constant $C$ whereas frame (b) indicates ZVC for various values of oblateness coefficients $A_1$ and $A_2$. For example, in frame (a) curves are labeled as $C_i, i=1,2,3,4$ for different values of Jacobi constant $ C_1=3.5, C_2=2.5, C_3=1.7$ and $C_4=1.0$ respectively whereas for in frame (b) $A_{11} (A_1=0.0, A_2=0.0), A_{12} (A_1=0.0025, A_2=0.0025)$, and $A_{13} (A_1=0.005,A_2=0.005)$ respectively. It is clear from frame (a), when $C$ is very large then the three primary bodies are separated with each other where the particle cannot move from one region to another. Again, when the values of  $C$ are small, connections open at two points where motion is possible and the body can never escape from the system. Further, we take $C$ even smaller then all the possible connections are opened i.e. inner and outer regions are opened and the particle can freely move from one allowed regions to another allowed region. On the other hand in frame (b), for increasing values of oblateness coefficients $A_1$ and $A_2$ respectively, their corresponding possible boundary regions increase where the particle can freely move from one side to another side. Therefore, we say that possible boundary region depends on the Jacobi constant as well as oblateness coefficients and observed that  how does the connection open for decreasing values of Jacobi constant  and increasing values of oblateness coefficients $A_1$ and $A_2$ respectively with other fixed values of the parameters.
 \section{Equilibrium points}
\label{sec:equn}
The coordinates of equilibrium points of the problem are obtained by equating R.H.S. of (\ref{eq:ux}) and (\ref{eq:vx}) to zero i.e. $ \Omega_{x} = \Omega_{y} = 0 $. In other words
\begin{eqnarray}
 && n^2x-\frac{(1-2\mu)(x-\sqrt{3}\mu)}{r_1^3}-\frac{3A_1(1-2\mu)(x-\sqrt{3}\mu)}{2r_1^5}\nonumber\\&&-\frac{(x+\frac{\sqrt{3}}{2}(1-2\mu))\mu}{r_2^3}-\frac{3A_2(x+\frac{\sqrt{3}}{2}(1-2\mu))\mu}{2r_2^5}\nonumber\\&&-\frac{(x+\frac{\sqrt{3}}{2}(1-2\mu))\mu}{r_3^3}=0,
 \label{eq:eqpt}
\end{eqnarray}
and
\begin{eqnarray}
&&\ n^2y-\frac{(1-2\mu)y}{r_1^3}-\frac{3A_1(1-2\mu)y}{2r_1^5}-\frac{(y-\frac{1}{2})\mu}{r_2^3}\nonumber\\&&-\frac{3A_2(y-\frac{1}{2})\mu}{2r_2^5}-\frac{(y+\frac{1}{2})\mu}{r_3^3}=0. \label{eq:eqpt1}
\end{eqnarray}
Solving  above equations for $\mu=0.2$ and different values of oblateness coefficient $ A_1$ and $A_2$, we obtain two collinear $L_{1,2}$ points on the $x$-axis and six non-collinear equilibrium points $L_i, i=3,4,...,8$ depicted in Figs. \ref{fig:fig8} and \ref{fig:fig9}.
\subsection{Equilibrium points when $y=0$}
The equilibrium points at $x$-axis are the solutions of Eqs. (\ref{eq:eqpt}) and (\ref{eq:eqpt1}) when $y = 0$, which give
\begin{eqnarray}
 &&f(x,0)= n^2x-\frac{(1-2\mu)(x-\sqrt{3}\mu)}{|x -\sqrt{3}\mu|^3}\nonumber\\&&-\frac{3A_1(1-2\mu)(x-\sqrt{3}\mu)}{2|x -\sqrt{3}\mu|^5}-\frac{2(x+\frac{\sqrt{3}}{2}(1-2\mu))\mu}{\left((x +\frac{\sqrt{3}}{2}(1-2 \mu))^2+\frac{1}{4}\right)^{\frac{3}{2}}}\nonumber\\&&-\frac{3A_2(x+\frac{\sqrt{3}}{2}(1-2\mu))\mu}{2\left((x +\frac{\sqrt{3}}{2}(1-2 \mu))^2+\frac{1}{4}\right)^{\frac{5}{2}}}=0.
 \label{eq:coln}
\end{eqnarray}
Now, solving the above expression using initial conditions, we get equilibrium points for various values of the oblateness coefficients. We observed that it has only two real roots and other are complex conjugates. Also, we noticed that for fixed values of at  $A_2=0.0015$ and for increasing values of $A_1 (0<A_1<1)$, equilibrium points at $x$-axis shifted from left to right, whereas for fixed values of $A_1=0.0015$ and for increasing value of $A_2 (0<A_2<1)$, equilibrium point $L_1$ shifted form left to right while  $L_2$ point is shifted form right to left which are shown in Table \ref{tab:t1}.
\begin{table}[h]
\scriptsize
\caption{Equilibrium points at $x$-axis \label{tab:t1}}
\begin{tabular}{@{}crrr@{}}
\tableline
$A_2=0.0015$\\$A_1$ & $L_1$ &$L_2$  \\
\tableline
0.0000&-0.953071&1.122780\\
0.0015&-0.952215&1.123100\\
0.0030&-0.951362&1.123420\\
0.0045&-0.950511&1.123730\\
0.0060&-0.949662&1.124040\\
0.0075&-0.948816&1.124350 \\
\tableline
$A_1=0.0015$\\$A_2$ & $L_1$ &$L_2$ \\
\tableline
0.0000&-0.952525&1.123770\\
0.0015&-0.952215&1.123100\\
0.0030&-0.951908&1.122440\\
0.0045&-0.951602&1.121770\\
0.0060&-0.951299&1.121110\\
0.0075&-0.950997&1.120460\\
\tableline
\end{tabular}
\end{table}

\begin{figure}
 \plotone{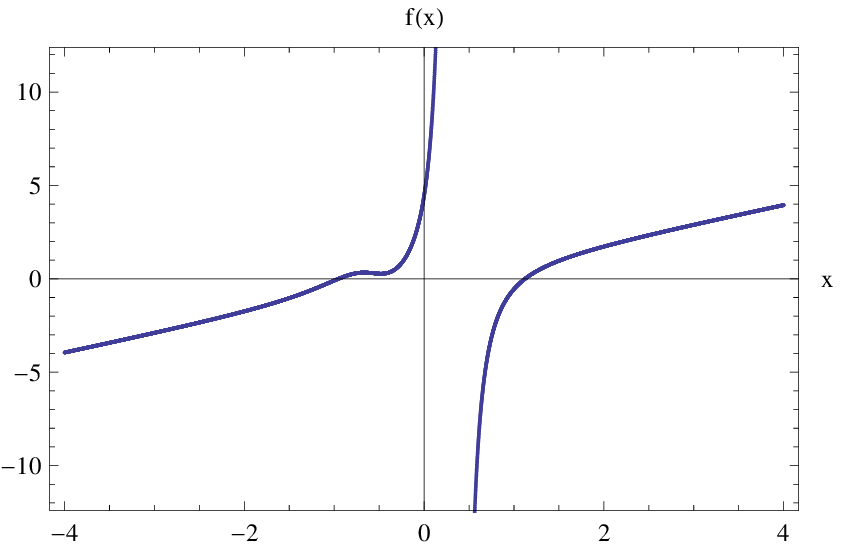}
\caption{Equilibrium point at $y=0$ for $\mu=0.2, A_1=0.0$ and $A_2=0.0015$.}\label{fig:fig7}
\end{figure}

We plot graph of equation (\ref{eq:eqpt}) when $y=0$ and fixed values of parameters $\mu=0.2, A_1=0.0$ and $A_2=0.0015$. From Fig. \ref{fig:fig7}, we observe that it intersect at only two points i.e. at $L_1=-0.953071$ and $L_2=1.122780$. From this figure as well as numerical computation we see that system has only two real roots and others are complex conjugates. Also, for other values of $A_1$ and $A_2$, number of equilibrium points remain same.

\subsection{Non-collinear points}
The non-collinear points are the solutions of Eqs. (\ref{eq:eqpt}) and (\ref{eq:eqpt1}) when $y \ne 0$, which gives
\begin{eqnarray}
 &&f(x,y)= n^2x-\frac{(1-2\mu)(x-\sqrt{3}\mu)}{r_1^3}\nonumber\\&&-\frac{(x+\frac{\sqrt{3}}{2}(1-2\mu))\mu}{r_2^3}-\frac{3A_1(1-2\mu)(x-\sqrt{3}\mu)}{2r_1^5}\nonumber\\&&-\frac{3A_2(x+\frac{\sqrt{3}}{2}(1-2\mu))\mu}{2r_2^5}\nonumber\\&&-\frac{(x+\frac{\sqrt{3}}{2}(1-2\mu))\mu}{r_3^3}=0,
 \label{eq:tript}
\end{eqnarray}
and
\begin{eqnarray}
&& g(x,y)= n^2y-\frac{(1-2\mu)y}{r_1^3}-\frac{3A_1(1-2\mu)y}{2r_1^5}\nonumber\\&&-\frac{(y-\frac{1}{2})\mu}{r_2^3}-\frac{3A_2(y-\frac{1}{2})\mu}{2r_2^5}\nonumber\\&&-\frac{(y+\frac{1}{2})\mu}{r_3^3}=0. \label{eq:tript1}
\end{eqnarray}
Solving equation (\ref{eq:tript}) and (\ref{eq:tript1}), we get non-collinear equilibrium points for different values of the oblateness coefficients $A_1$ and $A_2$ respectively.
For fixed value of $A_2$ and increasing values of $A_1$ as well as for fixed $A_1$ and increasing values of $A_2$, co-ordinates of non-collinear points $L_i, i=3, 4, ..., 8$ increase or decrease which are shown in Table \ref{tab:t2}. 

When the dominant primary bodies are oblate spheroids then we observe that as the oblateness coefficient $A_2$ increases from $0.0$ to $0.6$ for fixed value of $A_1=0.0015$, number of equilibrium points are eight but when $A_2$ increases from $0.7$ to $0.9$, the problem has then seven equilibrium points because $L_3$ approaches to $L_8$ point. Also, when $A_1=0.0$ and $A_2=1.0$ then the non-collinear equilibrium points $L_3$ and $L_8$ coincide on the collinear point $L_1$ and in consequence problem  has six equilibrium points. However, when $A_1=1.0$ and $A_2=0.8$ then equilibrium points become seven since $L_4$ reaches $L_8$ point, whereas oblateness coefficient $A_1$ increases form $0.0$ to $0.9$ for fixed value of $A_2=0.0015$, number of equilibrium points remain eight.  Further, we noticed that when $\mu=0.005$ and $A_1=A_2=0$ then our results agree with the results of \citep{Papadouris2013Ap&SS.344...21P}, their configuration was the mirror image of our configuration as depicted in Fig. \ref{fig:fig10}.
\begin{figure}
 \plotone{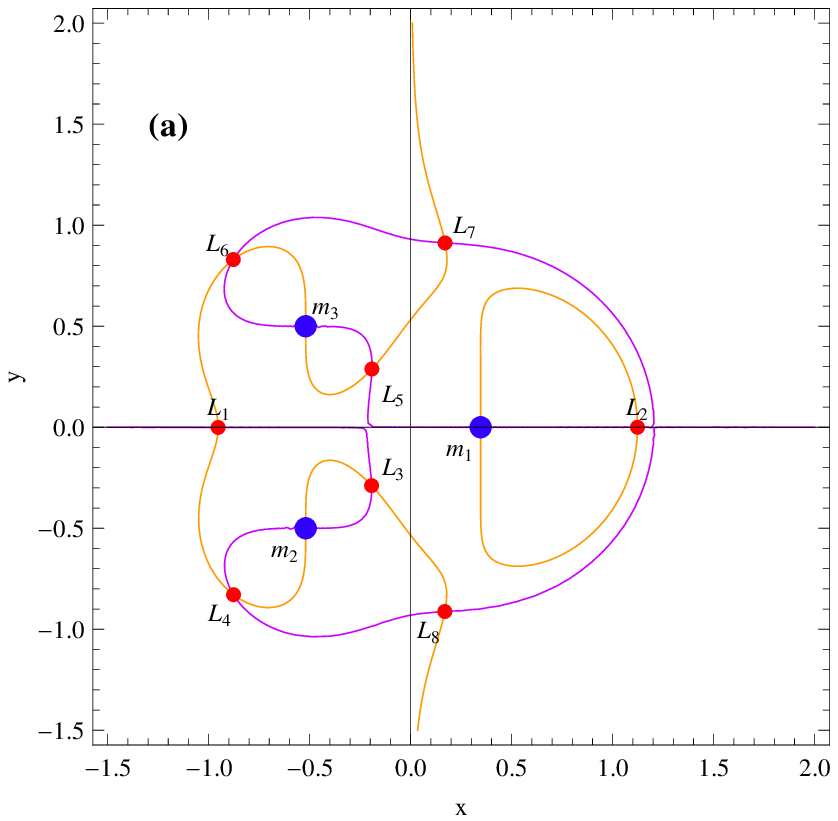}
\caption{The eight equilibrium points for $\mu=0.2, A_1=0.0$, and  $A_2=0.0015$.}\label{fig:fig8}
\end{figure}
\begin{figure}
 \plotone{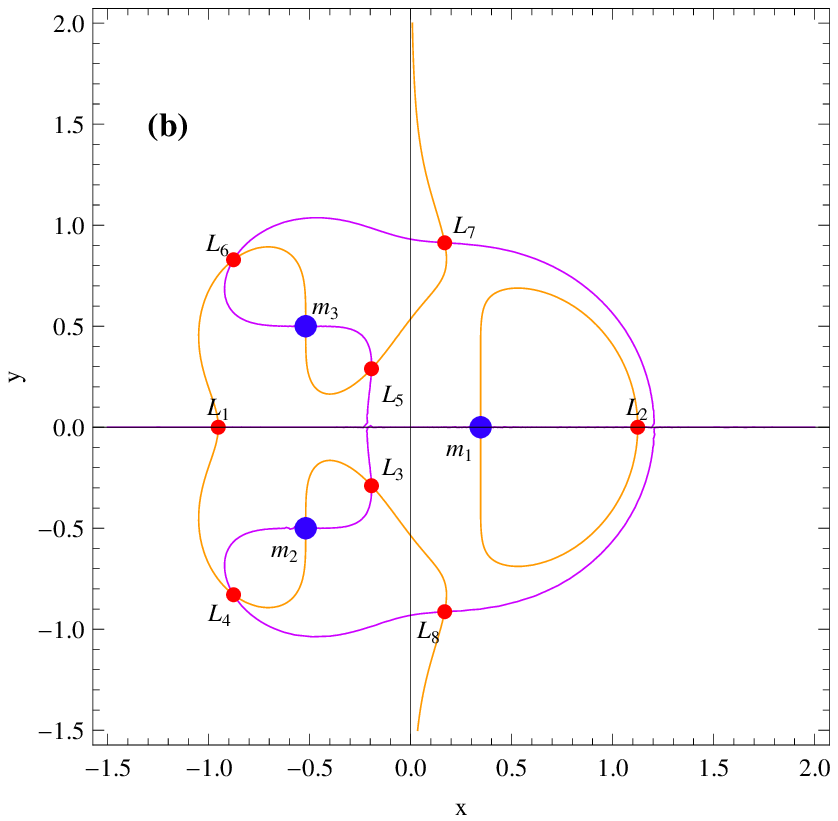}
\caption{The eight equilibrium points for $\mu=0.2, A_1=0.0015$ and $A_2=0.0$.}\label{fig:fig9}
\end{figure}
\begin{figure}
 \plotone{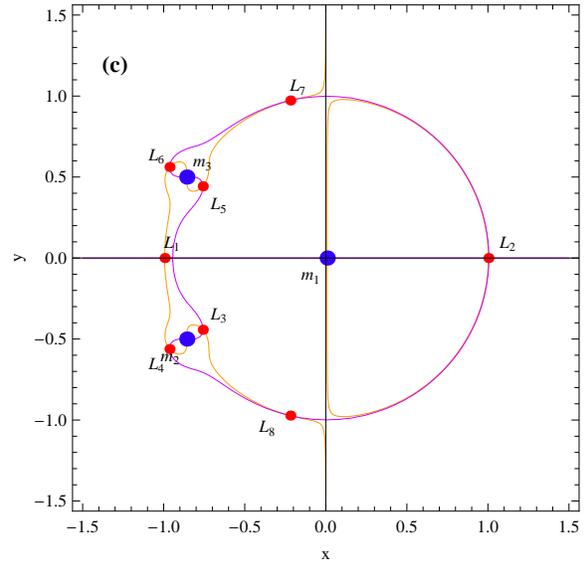}
\caption{The eight equilibrium points for $\mu=0.005, A_1=0.0$ and $A_2=0.0$.}\label{fig:fig10}
\end{figure}
\begin{figure}
 \plotone{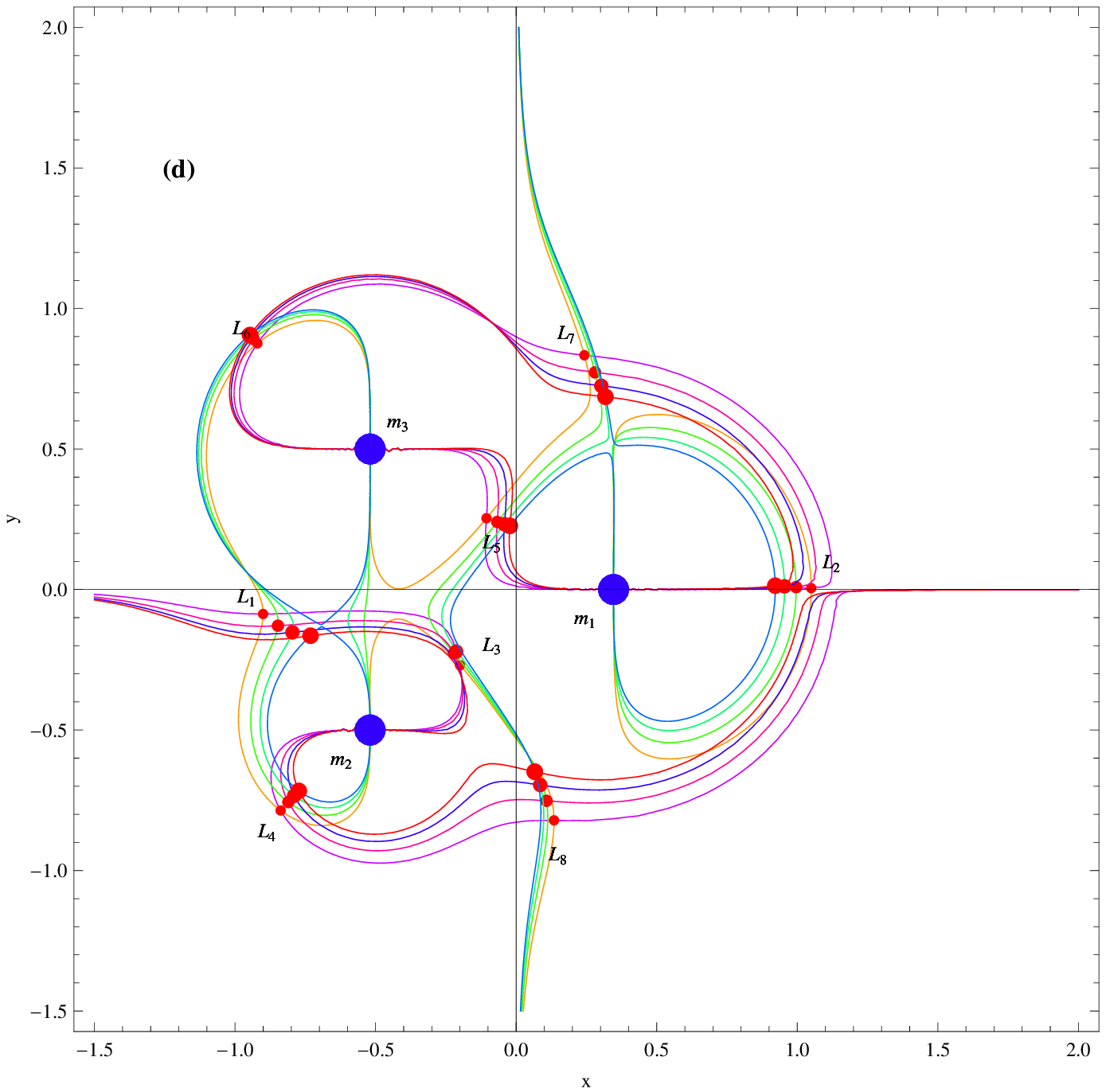}
\caption{The eight equilibrium points for $\mu=0.2, A_1=0.0015$ and increasing value of  $A_2 ~(0<A_2<1)$.}\label{fig:fig11}
\end{figure}
\begin{figure}
 \plotone{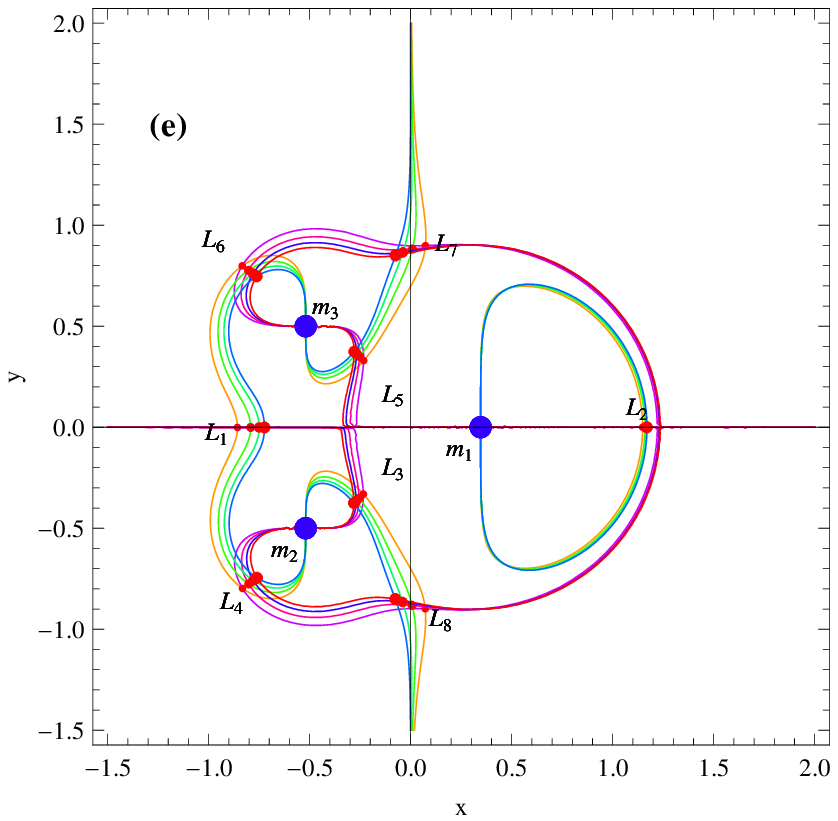}
\caption{The eight equilibrium points for $\mu=0.2, A_2=0.0015$ and increasing value of  $A_1~ (0<A_1<1)$}\label{fig:fig12}
\end{figure}

For fixed $A_1=0.0, A_2=0.0015$ as well as $A_1=0.0015, A_2=0.0$, we observe that second and third primary bodies form dumbell shape of the curve( Figs. \ref{fig:fig8} and \ref{fig:fig9}). However, the lower loop of the third primary body is disconnected, whereas one of the loop of second primary body reduces due to an increase in value of $A_2$ for fixed $A_1$ (Fig. \ref{fig:fig11}). On the other hand, the dumbell shape of the second and third primary bodies are less affected due to increasing value of $A_1$ for fixed value of $A_2$ (Fig. \ref{fig:fig12}). In Figs. \ref{fig:fig11} and \ref{fig:fig12}, we have used size of point to show the shifting of equilibrium points i.e. the equilibrium points shifted towards the large point size or along with increasing pointsize due to presence of oblateness coefficients. For $A_1=0.0015$ and $A_2 ~(0<A_2<1)$, $L_1, L_4$ and $L_8$ are attracted to second primary body, whereas  $L_2, L_5$ and $L_7$ are attracted towards the first primary body and it happens due to the attraction of the oblate bulge. Also, we see that $L_3$ and $L_6$ have very less effect of the parameters (Fig. \ref{fig:fig11}). Further, for $A_2=0.0015$ and  $A_1~ (0<A_1<1)$, $L_3, L_4$ and $ L_8$ are attracted towards the second primary body while $L_5, L_6$ and $L_7$ are attracted towards the third primary. Moreover, $L_2$ has very less effect of the parameters but $L_1$ is attracted by the first primary body due to same mass parameter values of second and third primary as shown in Fig. \ref{fig:fig12}.

\section{Linear stability of non-collinear points}
\label{sec:lsncp}
To analyze the possible motions of the infinitesimal body in a small displacement of the equilibrium points $(x_0, y_0)$, we first make infinitesimal change  $\xi $ and $\eta$ in its coordinates i.e. $x=x_0+\xi$ and $y=y_0+\eta$  such that the displacement becomes
 \begin{eqnarray}
  \xi=P e^{\lambda t},\quad \eta=Q e^{\lambda t}, \label{lstd1}
 \end{eqnarray}
where $P$, $Q$ are constants and $\lambda$ is parameter. Substituting these values into equations (\ref{eq:ux}) and (\ref{eq:vx}), we get differential equations of second order in $\xi $ and $\eta$ respectively \citep{murray1999solar}
 \begin{eqnarray}
 && \ddot \xi-2 n \dot \eta= \xi  \Omega^0_{xx}+\eta \Omega^0_{xy},\nonumber\\&&
  \ddot \eta+2 n \dot \xi= \xi  \Omega^0_{yx}+\eta \Omega^0_{yy}, \label{eq:varxieta}
 \end{eqnarray}
 where superfix $0$ indicates that the values are computed at the equilibrium point $(x_0, y_0)$. Again, substituting $\xi=P e^{\lambda t},\quad \eta=Q e^{\lambda t}$ in equation (\ref{eq:varxieta}) and simplifying, we obtain
  \begin{eqnarray}
 (\lambda^2-\Omega^0_{xx})P+(-2 n \lambda-\Omega^0_{xy})Q=0,\label{eq:pq1} \\
 (2 n \lambda-\Omega^0_{yx})P+(\lambda^2-\Omega^0_{yy})Q=0. \label{eq:pq2}
 \end{eqnarray}
 Now, the condition of nontrivial solution is that the determinant of the coefficients matrix of the above system should be zero i.e.
 \[
 \begin{vmatrix}
 \lambda^2-\Omega^0_{xx}& -2 n \lambda-\Omega^0_{xy} \\2 n \lambda-\Omega^0_{yx}& \lambda^2-\Omega^0_{yy}\\
 \end{vmatrix}
 =0.
 \]
 Therefore, from above matrix we obtain a quadratic equation in $\lambda^{2}$ known as characteristic equation:
\begin{eqnarray}
 &&\lambda^4+(4 n^2-\Omega^0_{xx}-\Omega^0_{yy}){\lambda^2}+\nonumber\\&&{(\Omega^0_{xx}}{\Omega^0_{yy}}-{\Omega^0}^2_{xy})=0.  \label{eq:ce} \end{eqnarray}
The four roots of characteristic equation (\ref{eq:ce}) play a crucial role to determine the orbits of equilibrium points. An equilibrium point will be stable if the above equation evaluated at the equilibrium, has four pure imaginary roots or complex roots with negative real parts. This happens if the following conditions 
 \begin{eqnarray}
 &&(4n^2-\Omega^0_{xx}-\Omega^0_{yy})^2-4(\Omega^0_{xx}\Omega^0_{yy}-(\Omega^0_{xy})^2)>0,\nonumber \\&&
 (4n^2-\Omega^0_{xx}-\Omega^0_{yy})>0,\nonumber\\&&
 \Omega^0_{xx}\Omega^0_{yy}-(\Omega^0_{xy})^2>0,
 \end{eqnarray}
 are satisfied simultaneously.

Now, using the determinant of the characteristic equation(\ref{eq:ce}) we obtain
\begin{eqnarray}
 &&(4.1407+14.8725A_1-15.0645A_2)\nonumber\\&&-(30.2203+101.6660A_1+31.8127A_2)\mu\nonumber\\&&-(191.3510+951.4380A_1+244.7510A_2)\mu^2>0, \label{eq:detmu}
\end{eqnarray}
which is a quadratic equation in $\mu$. Therefore, its root are given as
\begin{eqnarray}
&& \mu_1=\frac{s_1}{2(191.351+951.438A_1+244.751A_2)},\nonumber\\&&
\mu_2=\frac{s_2}{2(191.351+951.438A_1+244.751A_2)},\nonumber
\end{eqnarray}
where
\begin{eqnarray}
\nonumber&&s_{1,2}=-(30.2203+101.6660A_1+31.8127A_2)\\&&\mp\sqrt{(4082.61+33286.90A_1-5553.86A_2)}.\nonumber
\end{eqnarray}
These roots satisfy condition (\ref{eq:detmu}) if either (i) $\mu-\mu_1>0$ and $\mu-\mu_2>0$ or (ii) $\mu-\mu_1<0$ and $\mu-\mu_2<0$, which implies that $\mu>max(\mu_1, \mu_2)$ and $\mu<min(\mu_1,\mu_2)$ and therefore roots lie in between $\mu_1<\mu<\mu_2$. For numerical results we use $x_7= 0.165510,~ y_7=0.912095,~ 0<A_1<1$, and $ 0<A_2<1$ then we obtain $\mu_1=-0.241421$ and $\mu_2=0.0874975$.

The linear stability of the Lagrange central configuration is very important in celestial mechanics and is defined by the inequality \citep{Gascheau1843, Routh1875PLMS....6...86R, Papadouris2013Ap&SS.344...21P}
 \begin{eqnarray}
  \frac{m_1 m_2+m_2m_3+m_3m_1}{(m_1+m_2+m_3)^2}<\frac{1}{27}, \label{eq:lsb}
\end{eqnarray}
where $m_1, m_2$ and $m_3$ are masses of the three primaries body.
As we assumed $m_3 \le m_2$ and the left term of equation (\ref{eq:lsb}) inequality is monotonically increasing in $m_3,\quad \forall  m_3 \in (0,m_2)$, with maximum at $m_3 = m_2$. Therefore the stability condition becomes $-81m_2^2+54m_2<1,\quad \forall  m_2$, consequently we get $\frac{1}{9}(3-2\sqrt{2})\le m_2 \le \frac{1}{9}(3+2\sqrt{2})$. From this inequality we obtain mass parameter as  $0.0190637 \le \mu \le 0.647603$. From (\ref{eq:detmu}) we get two values of mass parameter out of which one value lies within above interval of $\mu$ which shows that non-collinear points are stable. 

\begin{figure}[h]
 \plotone{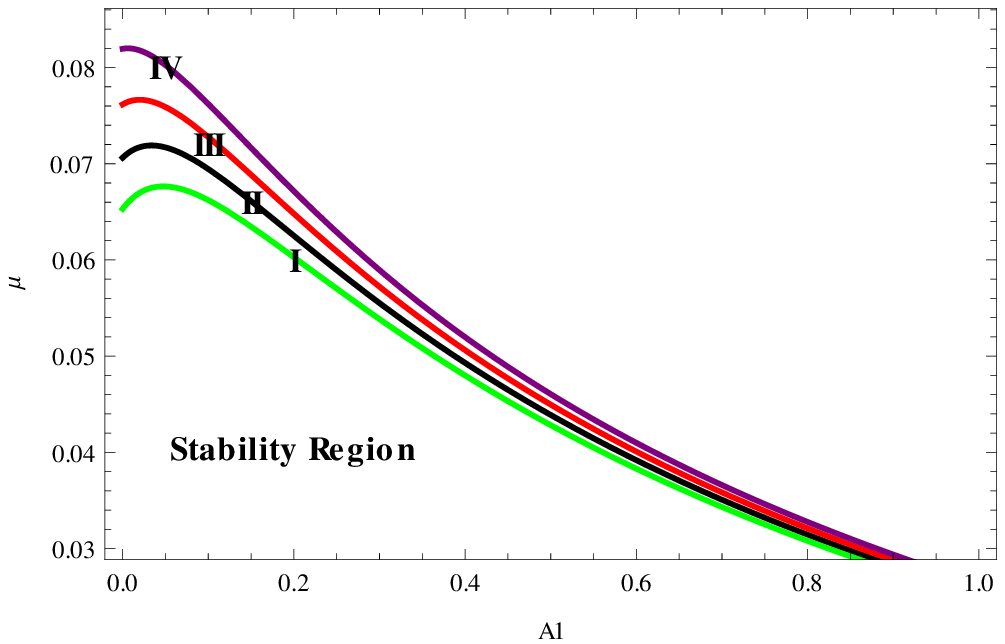}
\caption{Stability region of non-collinear points for fixed values of (I) $A_2=0.08$ (II)$A_2=0.06$ (III)$A_2=0.04$ (IV)$A_2=0.02$, varying $0<A_1<1.$ \label{fig:fig6}}
\end{figure}
In Fig.\ref{fig:fig6}, we have depicted the graph $A_1$ verses $\mu$ for different fixed values of $A_2$ and it is observed that for increasing values of $A_1$ and $A_2$, value of $\mu$ decreases consequently stability region decreases monotonically.
\section{Poincar\'{e} surfaces of section}
\label{sec:pss}
In the restricted four-body problem, Poincar\'{e} surface of section is very useful for finding stable periodic and quasi-periodic orbits around the primaries.
In order to determine Poincar\'{e} surface of section (PSS) of the infinitesimal body at any instant, it is necessary to know its position $(x, y)$ and velocity $(\dot x,\dot y)$, which correspond to a point in a four dimensional phase space. We have constructed surface of section on the $ x \dot x$-plane by taking $ y = \dot x = 0$ and $\dot y >0$ with the help of Event Locator Method of Mathematica$^{\textregistered}$\cite{wolfram2003mathematica}. 
This is a good technique to determine the regular or chaotic nature of the trajectory. On the other hand, if there are smooth, well defined islands, then the behavior of the  trajectory is likely to be regular. Whereas, if the curves shrink down to a point, it represents a periodic orbit. Apart form that, we have obtained PSS at the values of Jacobi constant $C$ for a certain values of $x$ and $\dot x $ while each orbit is determined with initial conditions:
\begin{eqnarray}
 && x=x_0,\quad y=0,\quad \dot x=0,\nonumber\\&&
\dot y =\sqrt{b_1+ n^2 x_0^2-\dot x_0^2-C},
\end{eqnarray}
where
\begin{eqnarray}
&& \nonumber b_1=\frac{2(1-2\mu)}{(x -\sqrt{3}\mu)}+\frac{3\mu}{\left((x +\frac{\sqrt{3}}{2}(1-2 \mu))^2+\frac{1}{4}\right)^{\frac{1}{2}}}\nonumber\\&&+\frac{(1-2\mu){A_1}}{(x -\sqrt{3}\mu)^{\frac{3}{2}}}+\frac{\mu{A_2}}{\left((x +\frac{\sqrt{3}}{2}(1-2 \mu))^2+\frac{1}{4}\right)^{\frac{3}{2}}}.\nonumber
\end{eqnarray}
Since in the above proposed system key quantities are the values of $C, A_1$ and $ A_2$ respectively. Therefore, we plot the graph of Poincar\'{e} surfaces of section for specific initial values $x_0=0.1, \ \dot x_0=0.3$, $y_0=-0.1$ with different values of Jacobi constant and oblateness coefficient respectively.
\begin{figure}[h]
 \begin{center}
  \plotone{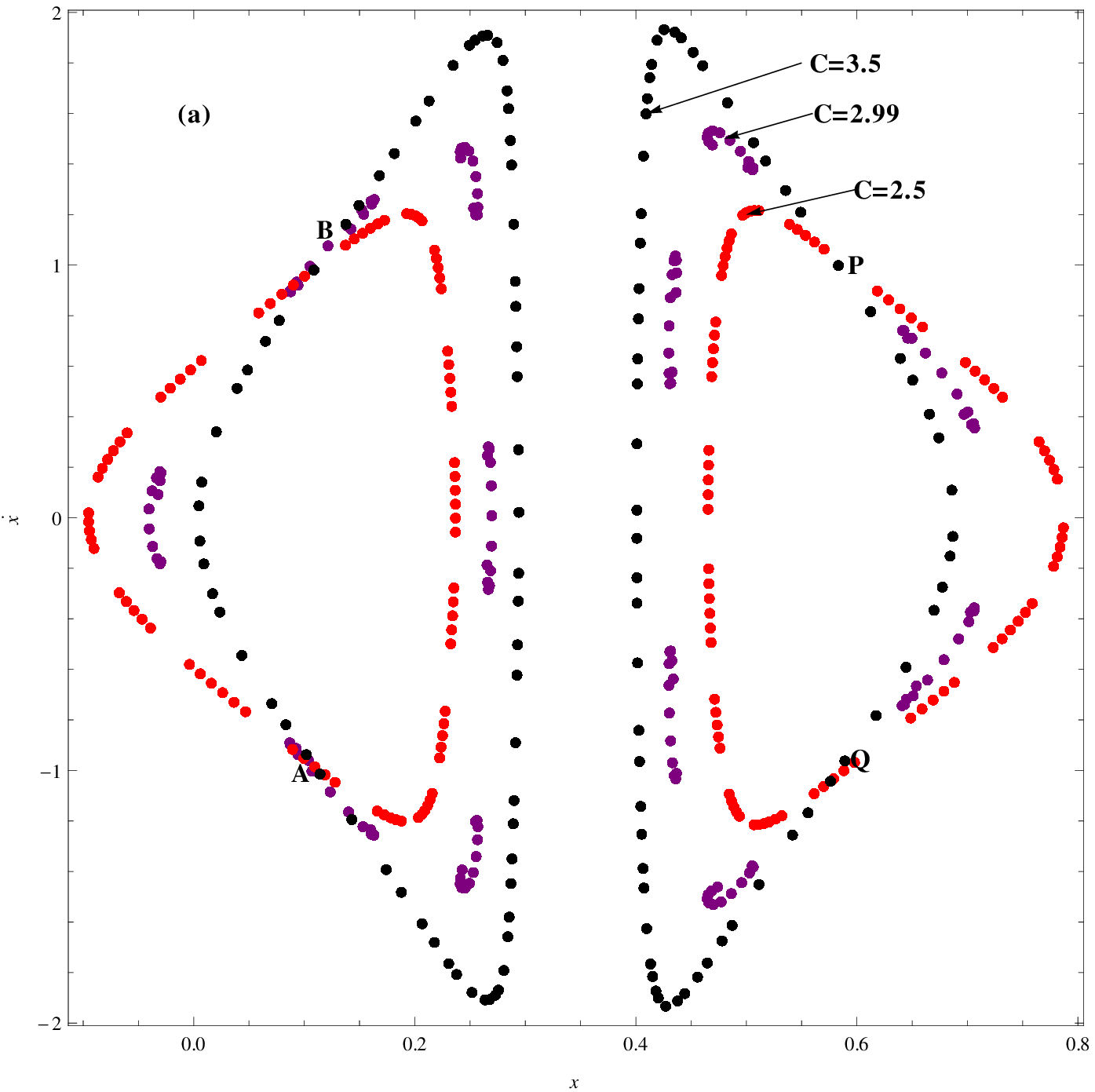}\plotone{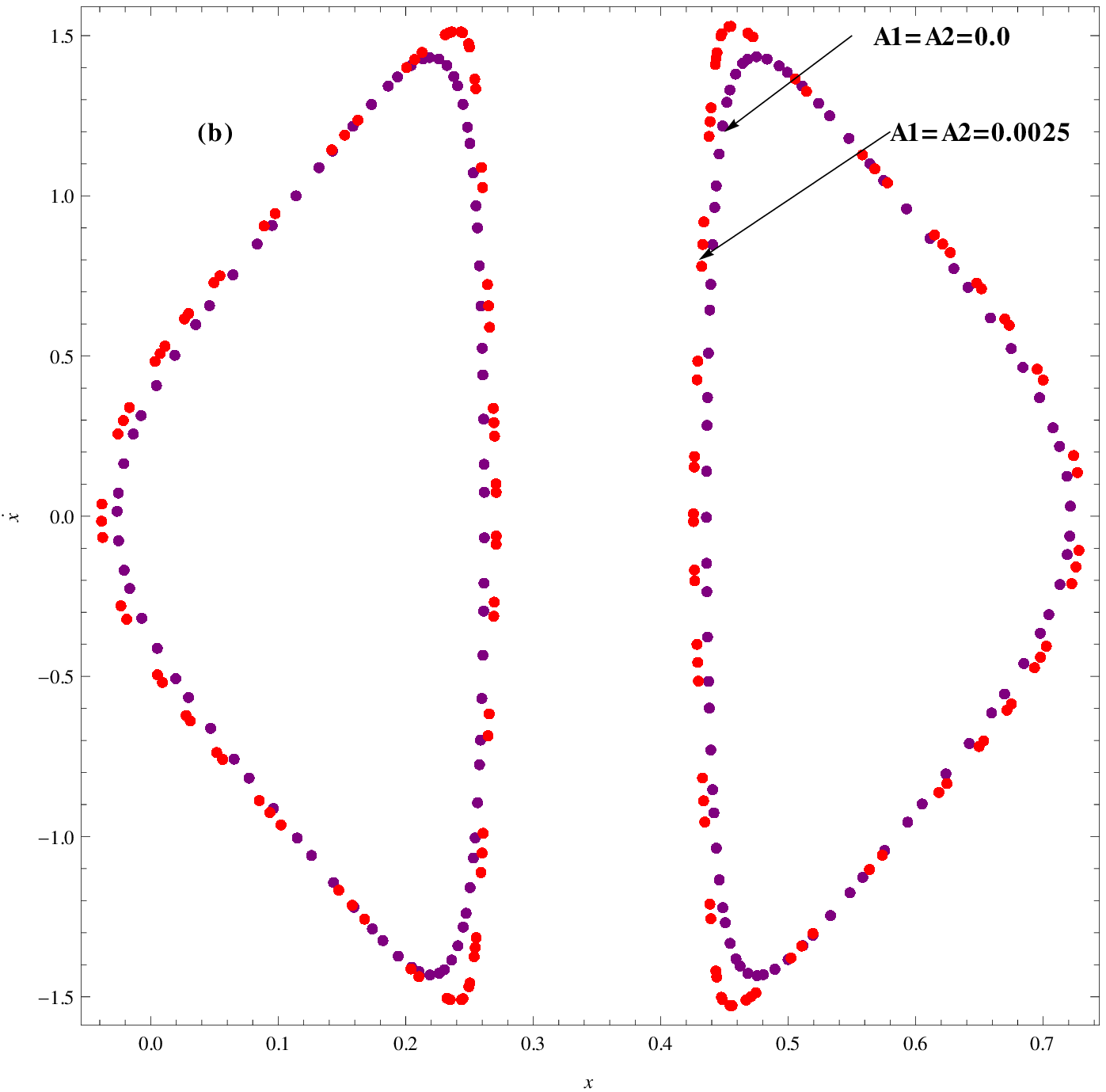}
\caption{Poincar\'{e} surface of section for the effect of Jacobi constant as well as oblateness.\label{fig:fig4}}
 \end{center}
\end{figure}
In Fig.\ref{fig:fig4}, we have shown two different characteristics of the system  i.e. the effect of oblateness coefficient as well as Jacobi constant. It is clear that a trajectory originated from  the neighborhood of equilibrium point, crosses Poincar\'{e} surfaces of section in bounded region and remains in that region for long time, which shows that the orbit about equilibrium point is stable. However, for various values of parameters, the bounded region changes i.e. if we increase oblateness coefficients $A_1$ and $A_2$ respectively, then the region expands (as shown in frame \ref{fig:fig4}(b)). Similarly if we increase the values of $C$ i.e. $C=2.5,\ 2.99$ and $C=3.5$, then the bounded region spans (as shown in frame \ref{fig:fig4}(a)). 

For a particular values of initial conditions $x_0=0.1, \ \dot x_0=0.3$ and $y_0=-0.1$ and different values of $C$, we observe that near the points $A(0.0946,-0.9327),\ B(0.1186,1.063), P(0.5789,1.011)$ and $ Q(0.6029,-0.9678)$ respectively, trajectories look like as they touch each other which shows that orbit is stable around the neighborhood of the equilibrium point.
\section{Basins of attraction} 
\label{sec:basin}
We determine basins of attraction of the equilibrium points with the help of Newton-Raphson method, provided an initial point $(x,y)$ and the mass parameter $\mu$ as well as oblateness coefficient $A_1$ and $A_2$ respectively are given. 

It is a good technique to find the convergence of trajectory originated from neighborhood of an equilibrium point. We present basins of attraction of a fixed points, means that the set of points converge towards a fixed point under successive iterations of some transformation. The set of points $(x,y)$ that are created as follows: 
\begin{eqnarray}
&& \nonumber \Omega_{x}(x,y,\mu,A_1,A_2) =0,\\&&
 \Omega_{y}(x,y,\mu,A_1,A_2) = 0,\label{eq:basn}
\end{eqnarray}
from which we obtain the equilibrium points of the problem. The algorithm of our problem takes the form
\begin{eqnarray}
 &&\nonumber x^{(n)}=x^{(n-1)}-\frac{\Omega_{x} \Omega_{yy}-\Omega_{y}\Omega_{xy}}{\Omega_{yy}\Omega_{xx}-\Omega^2_{xy}}\arrowvert_{x^{(n-1)},y^{(n-1)}},\\&&
y^{(n)}=y^{(n-1)}+\frac{\Omega_{x} \Omega_{yx}-\Omega_{y}\Omega_{xx}}{\Omega_{yy}\Omega_{xx}-\Omega^2_{xy}}\arrowvert_{x^{(n-1)},y^{(n-1)}},\nonumber\\
\end{eqnarray}
where $x^n$ and $y^n$ are the values of $x$ and $y$ at the $n^{th}$ step of the Newton-Raphson method.

Now, if the starting point $(x,y)$ converges rapidly to a specific root of the algebraic equation (\ref{eq:basn}), then this point $(x,y)$ is a member of the basin of attraction of the specific root. The Newton-Raphson method stops when the resulting successive approximation converges to an attractor, the convergence being terminated when the repetition is happened. If the iteration diverges, then the process is terminated after 100 iterations. The regions of the basins of attraction are constructed by applying a dense grid of node points in the $xy$-plane as starting points for the iteration.

\begin{figure}[h]
\begin{center}
  \includegraphics[scale=0.37]{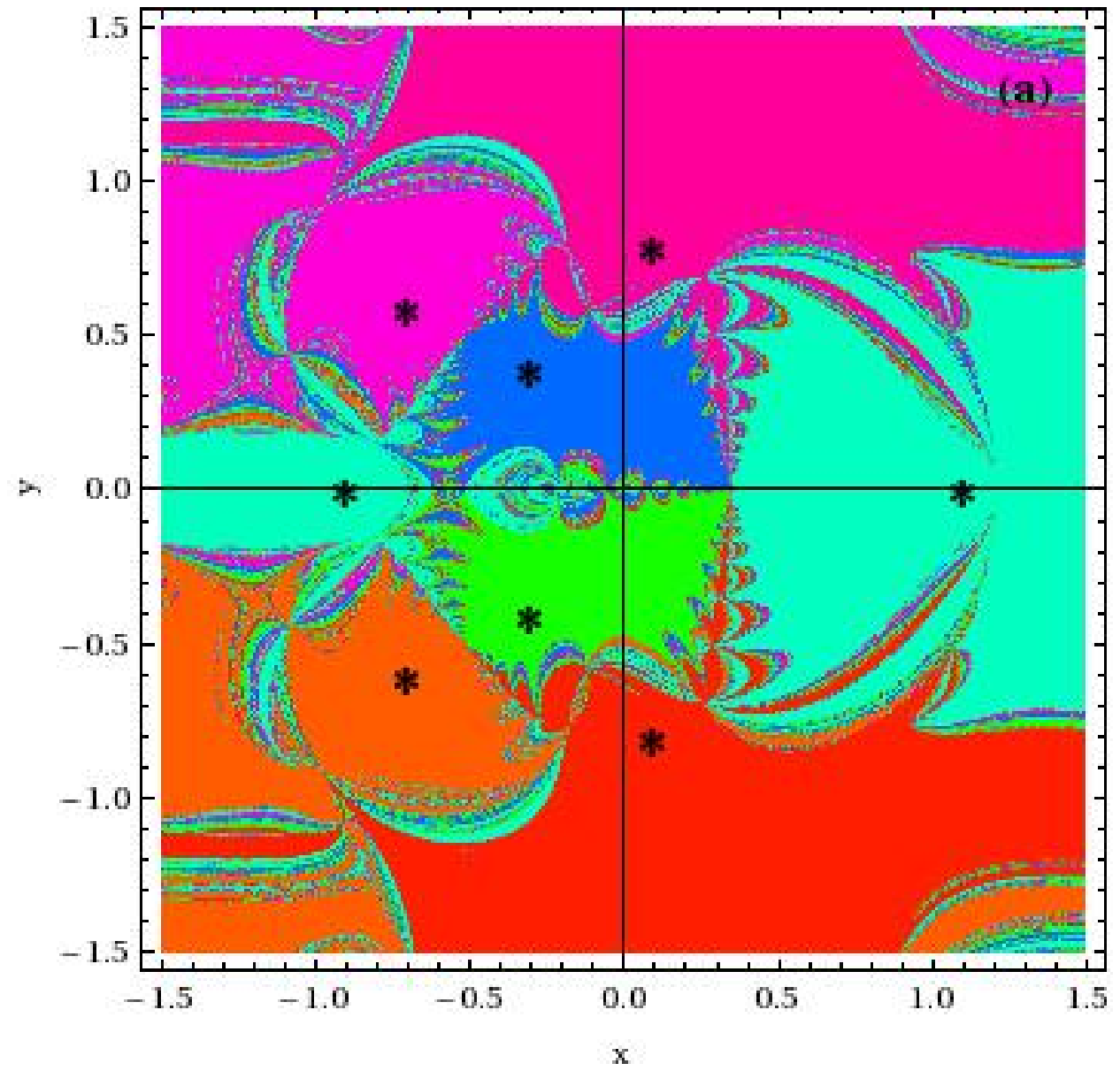}\includegraphics[scale=0.37]{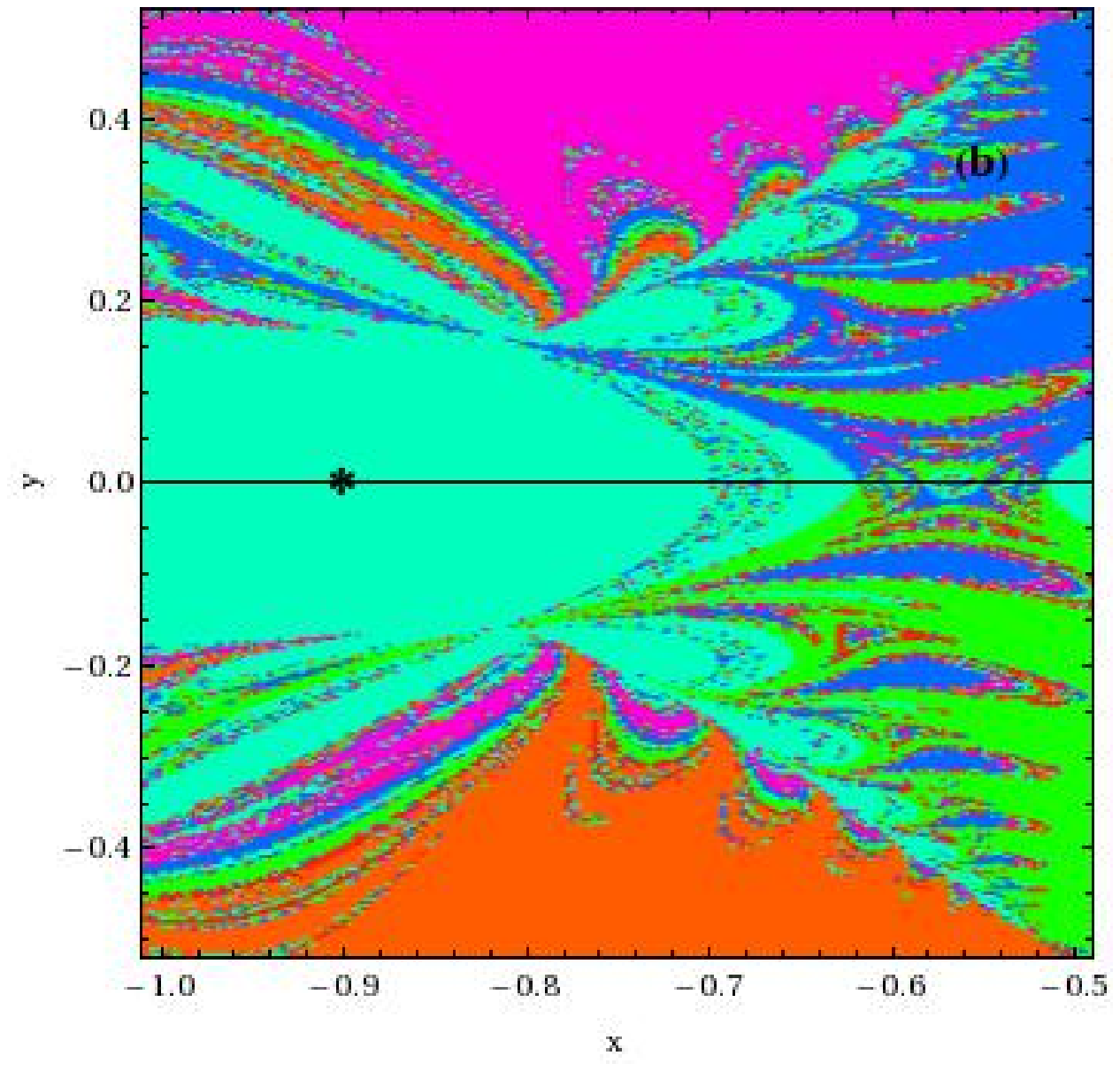}\\ \includegraphics[scale=0.37]{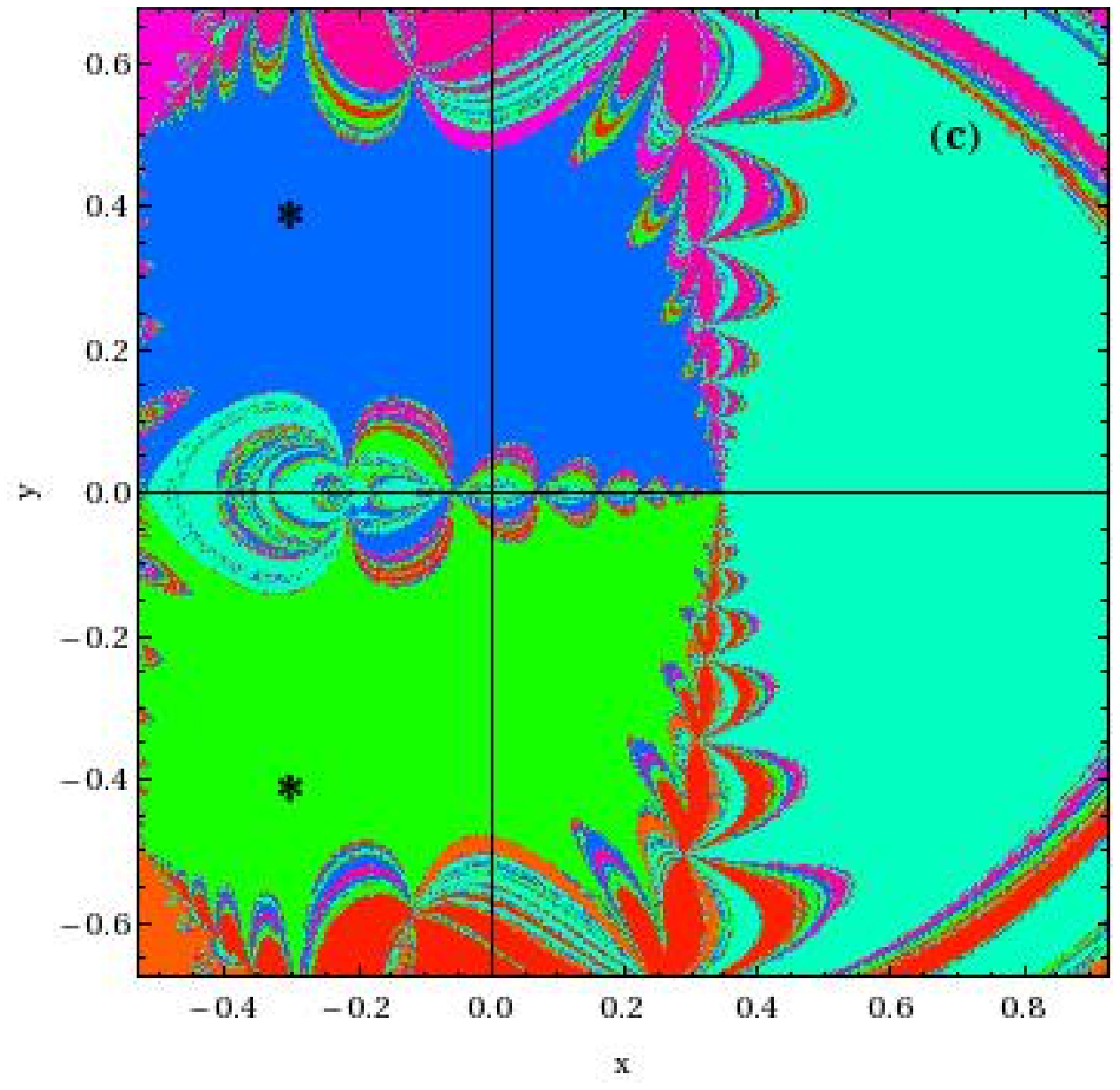}\includegraphics[scale=0.37]{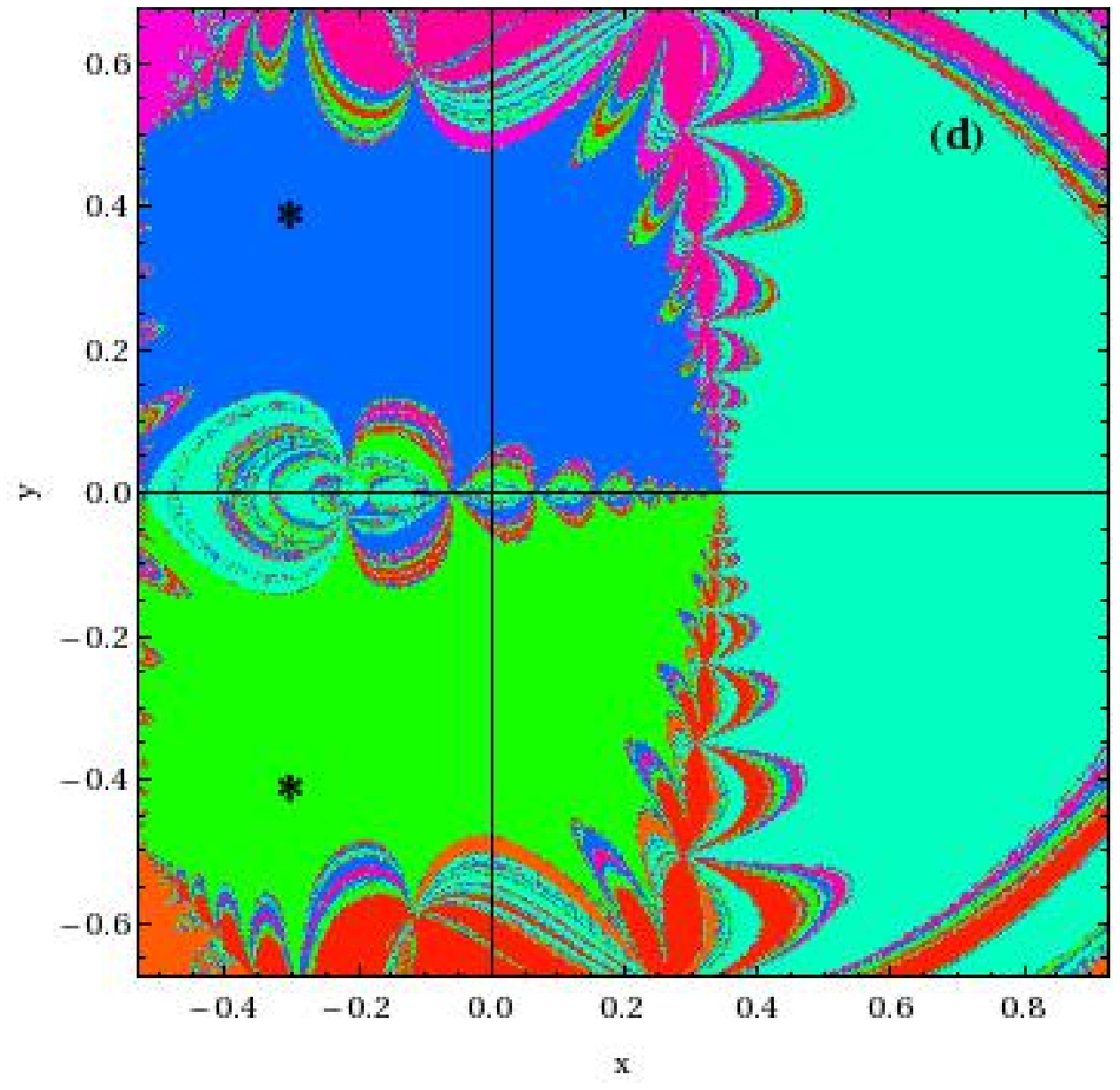}
\caption{(a) The regions of different colors denote the basins of attraction for the  equilibrium points except collinear points which are shown in the single color of the restricted four-body problem when oblateness coefficient $A_1=0.0015$ and $A_2=0.0015$ respectively. Whereas frames (b) and (d) show the zoom portion near the center of the frame (a). Frame (c) is zoom part of frame (a) when the oblateness coefficients are absence ($A_1=0.0$ and $A_2=0.0$). The positions of the eight attractors are indicated by small black stars.\label{fig:fig5}}
 \end{center}
\end{figure}

In Fig.\ref{fig:fig5}, we present the basins of attraction of the equilibrium points in the restricted four body problem which are shown in frame (a) whereas other frames are zoom portions of frame (a). 
For each basins of attraction we use different color and the equilibrium points are indicated by small stars. The existence of one very large body and other two small ones effects the structure of the basins substantially. The points of the attracting domain of the central zone are organized in diamond shaped parts, whose wavy sides have vague boundaries. Inside, these areas lie the equilibrium positions of that zone. The boundaries of the central part are not clearly defined. They look like a "chaotic sea". Again, outside the central zone the points of attractor is organized in mushroom shaped regions where the equilibrium points contain in this zone. The boundaries of the mushroom shaped regions are dispersed points. The dispersed points of this class are densely allocated on the boundaries of the dense areas of the attracting regions. In presence of oblateness coefficients $A_1$ and $A_2$, there is very less difference in absence of oblateness coefficients $A_1$ and $A_2$  respectively which are shown in frame (b) and frame (d). On the other hand, we can say that different combination of oblateness coefficient gives same nature of the problem. However, frame (c) indicates the zoom part of frame (a) when the oblateness coefficients are absent.
\section{Discussion and conclusion}
\label{sec:con}
We have studied restricted four-body problem (RFBP) introducing first two bigger primaries as oblate spheroids. The boundary regions for the motion of an infinitesimal body are obtained with the help of zero velocity surfaces at different values of Jacobi constant and fixed values of oblateness coefficients. We have found that the allowed possible regions of the motion of infinitesimal body decrease with increases values of the Jacobi Integral $C$. We have investigated orbit of the RFBP and found that in absence of oblateness coefficients, orbit looks like cote's spiral in the time interval $100\le t \le 200$,  whereas with effect of oblateness coefficient, orbit becomes regular when $0 \le t \le 200$.

We have determined the coordinates of equilibrium points at $y=0$ and non-collinear points at $y\ne0$, which depend on oblateness coefficient $A_1$ and $A_2$. We have noticed that for fixed value of $A_1=0.0015$ and increasing values of $A_2 (0<A_2<1)$ as well as for fixed value of $A_2=0.0015$ and increasing values of $A_1 (0<A_1<1)$, system at $y=0$ has only two real roots called collinear points, whereas at $y\ne 0$ it has six real roots called non-collinear points. The oblateness coefficients affect the existence of the equilibrium points of the problem in hand, since for $ A_1=0.0015$ and increasing value of $A_2$ from $0.7$ to $0.9$, $L_3$ disappears by coalescing at the $L_8$ and consequently the problem has seven equilibrium points.  However, when the oblateness coefficient $A_1$ increases from $0.0$ to $0.9$ for fixed value of $A_2=0.0015$, number of equilibrium points remains eight. Two collinear equilibrium points always exist for every value of the oblateness coefficient.

We have also found that  for $A_1=0.0015$ and $A_2 ~(0<A_2<1)$, $L_1, L_4$ and $L_8$ are attracted by second primary, whereas  $L_2, L_5$ and $L_7$ are attracted towards the first primary and this happens due to the attraction of the oblate bulge. Also, we have seen that $L_3$ and $L_6$ have very less effect of the parameters. Furthermore, for $A_2=0.0015$ and  $A_1~ (0<A_1<1)$, $L_3, L_4$ and $ L_8$ are attracted towards the second primary while $L_5, L_6$ and $L_7$ are attracted towards the third primary. The $L_2$ point have very less effect of the parameters but $L_1$ is attracted by the first primary body due to same mass parameter values of second and third primary bodies respectively.

The non-collinear points are stable if the mass parameter $\mu$ belongs to the interval ($0.0190637,~ 0.647603$). With the help of PSS, it is observed that the stability region of an equilibrium point  gets expanded from the center due to effect of oblateness coefficients and for a particular set of values of initial conditions $x_0=0.1, \dot x_0=0.3$ and $y_0=-0.1$, the trajectories touch each other at the points $A(0.0946,-0.9327), B(0.1186,1.063), P(0.5789,1.011)$ and $ Q(0.6029,-0.9678)$ respectively which represents that orbit are stable around the neighborhood of the equilibrium point. Further, we have presented basins of attraction for the equilibrium points with the help of Newton Raphson method. These basins of attraction are described in the $xy$-plane, showing the attractor of the Newton iteration. Due to the presence of oblateness coefficients, we have found that boundaries of the basins of attraction for the equilibra are not clearly defined which shows the chaotic nature. Also, we observed that there is very less difference in basins of attraction compare to absence of oblateness coefficients. Since it is difficult to obtain an exact boundaries of the equilibra of the restricted four-body problem \citep{Douskos2010Ap&SS.326..263D, Baltagiannis2011IJBC...21.2179B}, further work is needed in this regard. This work may be applicable to study the motion of a test particle in the Sun-Earth-Moon-spacecraft as well as Sun-Jupiter-Trojan-spacecraft system.  

\acknowledgements{We are thankful to IUCAA, Pune for partially financial support to visit library and to use computing facility. We are also thankful to Prof. Bhola Ishwar, B.R.A. Bihar University, Muzaffarpur (India) and Mr. Ashok Kumar Pal, ISM, Dhanbad (India) for their valuable suggestions during the preparation of the manuscript.}


\clearpage
\begin{deluxetable}{rrrrrrrrrr}
\tabletypesize{\scriptsize}
\rotate
\tablecaption{Non-collinear equilibrium points \label{tab:t2}}
\tablewidth{0pt}
\tablehead{\colhead{$A_2=0.0015$}\\
\colhead{$A_1$}& \colhead{$L_3$} & \colhead{$L_4$} & \colhead{$L_5$} & \colhead{$L_6$} & \colhead{$L_7$} & \colhead{$L_8$} 
}
\startdata
0.0000&(-0.193457,~ -0.288846)&(-0.876813,~ -0.828971)&(-0.191977,~ 0.288315)&(-0.877914,~ 0.830136)&(0.170043,~ 0.912386)&(0.168924,~ -0.912255)\\
0.0015&(-0.193948,~ -0.289374)&(-0.876402, ~ -0.82869)&(-0.192469, ~ 0.288840)&(-0.877506,~ 0.829856)&(0.169129,~ 0.912330)&(0.168010, ~ -0.912197)\\
0.0030&(-0.194434, ~-0.289896)&(-0.875994, ~-0.828410)&(-0.192956, ~0.289360)&(-0.877099, ~0.829577)&(0.168219, ~0.912273)&(0.167098, ~ -0.912139)\\
0.0045&(-0.194915, ~-0.290411)&(-0.875586, ~-0.828131)&(-0.193437, ~0.289873)&(-0.876693, ~0.829299)&(0.167312, ~0.912215)&(0.166191, ~-0.912079)\\
0.0060&(-0.195391, ~-0.290920)&(-0.875180, ~-0.827853)&(-0.193913, ~0.29038)&(-0.876289, ~0.829022)&(0.166409,~ 0.912156)&(0.165287,~ -0.912018)\\
0.0075&(-0.195861, ~-0.291423)&(-0.874776, ~-0.827576)&(-0.194385, ~0.290881)&(0.875886, ~0.828746)&(0.165510,~ 0.912095)&(0.164387,~ -0.911956) \\
 \hline\\
$A_1=0.0015$\\$A_2$\\\hline

0.0000&(-0.193927, ~-0.289496)&(-0.876758, ~-0.829082)&(-0.193927, ~0.289496)&(-0.876758, ~0.829082)&(0.168296, ~ 0.913002)&(0.168296, ~ -0.913002)\\
0.0015&(-0.193948, ~-0.289374)&(-0.876402, ~-0.828690)&(-0.192469, ~0.288840)&(-0.877506, ~0.829856)&(0.169129, ~0.912330)&(0.168010,~ -0.912197)\\
0.0030&(-0.193970, ~-0.289252)&(-0.876048, ~-0.828300)&(-0.191051, ~0.288205)&(-0.878237, ~0.830614)&(0.169956, ~0.911660)&(0.167724,~ -0.911395)\\
0.0045&(-0.193991, ~-0.289129)&(-0.875695, ~-0.827912)&(-0.189669, ~0.287588)&(-0.878951, ~0.831356)&(0.170776, ~0.910990)&(0.167438, ~ -0.910594)\\
0.0060&(0.194013, ~-0.289006)&(-0.875343, ~-0.827524)&(-0.188321, ~0.286988)&(-0.879650, ~0.832082)&(0.171590, ~0.910321)&(0.167153, ~ -0.909796)\\
0.0075&(-0.194035, ~-0.288883)&(-0.874992, ~0.827138)&(-0.187006, ~0.286405)&(-0.880334, ~0.832794)&(0.172397, ~0.909654)&(0.166869, ~ -0.908999)\\
\enddata
\end{deluxetable}

\end{document}